\documentclass[english,11pt]{article}
          \usepackage{babel}
          \usepackage[T1]{fontenc}
          \usepackage[latin2]{inputenc}
          \usepackage{pslatex}
\begin{document}

\title{\bf Curious Variables Experiment (CURVE). \\
Three Periodicities of BF Ara\footnote{Based on observations at SAAO}}
\author{A. ~O~l~e~c~h$^1$, ~A. R~u~t~k~o~w~s~k~i$^1$,
~~and~~ A. ~S~c~h~w~a~r~z~e~n~b~e~r~g~-~C~z~e~r~n~y$^{1,2}$}

\date{$^1$ Nicolaus Copernicus Astronomical Center,
Polish Academy of Sciences,
ul.~Bartycka~18, 00-716~Warszawa, Poland,\\
{\tt e-mail: (olech,rudy,alex)@camk.edu.pl}\\
~\\
$^2$ A. Mickiewicz University Observatory, ul. S{\l}oneczna 36,
60-286 Pozna\'n, Poland.
}

\maketitle

\begin{abstract}

We report CCD photometry of the dwarf nova BF Ara throughout fifteen
consecutive nights in quiescence. Light curve in this interval is
dominated by a large amplitude ($\sim 0.8$ mag) modulation consisting two
periods. Higher amplitude signal is characterized by period of 0.082159(4)
days, which was increasing at the rate of $\dot P/P_{\rm sh} =
3.8(3)\cdot 10^{-5}$. Weaker and stable signal has period of
0.084176(21) days. Knowing the superhump period of BF Ara determined by
Kato et al. (2003) and equal to 0.08797(1) days, the first modulation is
interpreted as quiescent negative superhump arising from retrograde
precesion of titled accretion disk and the latter one as an orbital
period of the binary. The respective period excess and defect are
$\epsilon_+ = 4.51\% \pm 0.03\%$ and $\epsilon_- = -2.44\% \pm 0.02\%$.
Thus BF Ara is yet another in-the-gap nova with
mass ratio of around $q\approx0.21$.

\end{abstract}

\maketitle

\section{Introduction}

The SU UMa variables are a subclass of dwarf novae characterized by the
presence of two types of eruptions - normal outburst having an amplitude
of 2-3 mag and repeating typically every few tens of days and
superoutburts with amplitude of 3-4 mag, lasting few times longer and
occuring every year or so.

These stars are belived to be binary systems consiting of the white dwarf
primary and low mass main-sequence secondary filling its Roche lobe and
loosing material which forms an accretion disc around the primary.
Almost all SU UMa stars are close binaries with orbital period from 1.25
to slightly over 2 hours.

This behaviour in now quite well understood within the frame of the
thermal-tidal instability model (see Osaki 1996 for review). Normal
outbursts are caused by the thermal instability connected with transition
of the material in the disc from neutral to ionized state. Superoutbursts
are a result of combined thermal and tidal instability, which, working
together, are very effective with sweeping out the matter from the disc.

During the superoutburst, the characteristic tooth-shape light modulations
with a period a few percent longer than the orbital period of the binary
are observed. They are most probably the result of accretion disc
"precession" (in fact it is not classical precession but change of the
position of the line of apsides) caused by gravitational perturbations
from the secondary. These perturbations are most effective when disc
particles moving in eccentric orbits enter the 3:1 resonance. Then the
superhump period is simply the beat period between orbital and
"precession" rate periods.

At the beginning of the 1990s, the quite simple class of SU UMas started
to be more complicated. The systems showing superhumps were divided into
four subgroups:

\begin{itemize}

\item WZ Sge stars, characterized by an extremely long quiescent state,
going into superoutburst every $\sim$10 years and showing no or very
infrequent ordinary outbursts (Patterson et al. 2002),

\item ordinary SU UMa stars,

\item ER UMa stars - systems characterized by an extremely short
supercycle (20-60 days), a short interval between normal outbursts
(3-4 days) and small amplitude (2--3 mag) of superoutbursts (Kato and
Kunjaya 1995, Robertson et al. 1995),

\item permanent superhumpers - high accretion rate systems being
permanently in superoutbursts (Skillman and Patterson 1993).

\end{itemize}
\smallskip

We now believe that the classes listed above represent
increasing mass transfer. Systems with mass transfer rates as low
as $10^{15}$ g/s are inactive, quiet and evolved stars of WZ Sge
type containing  brown dwarf degenerated secondary. Classical SU
UMa stars have mass transfer rates  one magnitude higer and go
into the superoutburt every year or so. ER UMa stars are high mass
transfer rate systems (few times per $10^{16}$ g/s) showing
frequent and relatively long superoutburst lasting for about half
of the supercycle. The explanation of the shortest supercycles
involves poorly understood mechanisms causing premature
quench of the eruption (Osaki 1995). And finally, permanent
superhumpers are system with accretion disc which is thermaly stable 
and tidaly unstable all the time, i.e. being in permanent state of
superoutburst.

\section{Observations and data reduction}

In order to firmly establish the mutual relations between these SU UMa
subclasses one has to study systems located near to the borders between
them. For these purposes we selected several poorly studied objects
which are very interesting from the point of view of the theory which
describes the origin of superhumps in dwarf novae systems. For a report
of our northern hemisphere work performed at the Ostrowik observatory see
previous reports on our CURVE project, e.g. Rutkowski et al. (2007).
Since a significant number of these systems is located in the southern
hemisphere, we applied for the observation time in South African
Astronomical Observatory (SAAO). For a pilot study we had been allocated
time from August 15 to September 11 on Elizabeth telescope at SAAO.

The observations were performed by one of us (A.R.) using a 1-m
Cassegrain telescope of 8.5-m focal length. The telescope worked with
CCD camera "STE3", back illuminated chip manufactured by SITe of size
$512\times 512$ pixels. The image scale was 0.31 arcsec/pix providing a
$158 \times 158$ arcsec field of view. The camera worked with liquid
nitrogen cooling. Technology of the whole system allows to minimize the
influence of noises like bias and dark current. Thanks to that it was
not necessary to obtain the bias and dark frames. The filters wheel was
equipped with Johnson-Cousins $UBV(RI)$ filters but most of observations
were obtained in white light as we wanted to examine temporal behaviour
of light curve and have good quality photometry of faint objects. We
were using exposure times ranging from 100 to 200 sec, depending on
weather conditions and actual brightness of the object. Data reduction
was performed using a standard procedure based on IRAF
package.\footnote{ IRAF is distributed by the National Optical Astronomy
Observatory, which is operated by the Association of Universities for
Research in Astronomy, Inc., under cooperative agreement with the
National Science Foundation.} The profile photometry has been derived
using the DAOphotII package (Stetson 1987). The differential photometry
provided measurements with accuracy in most cases around or below 0.01
mag.

\section{BF Ara}

One object from our southern sky survey is dwarf nova BF Ara. It
was discovered as a variable by Shapley and Swope (1934) but the
nature of its light variations remained unknown utill work of
Bruch (1983) who made $U$, $B$ and $V$ photometry of this object
during one night in its bright state. The light curve exhibited a
tooth-shape hump with an amplitude of 0.25 mag and period of
roughly two hours. The conlusion was that BF Ara might belong to
the SU UMa dwarf novae.

In the years 1997--2001 BF Ara was regulary monitored by VSNET
Collabolators. These data allowed Kato et al. (2001) to find that BF Ara
is one of the most active normal SU UMa stars with mean interval between
the succesive superoutbursts amounting to 83.4 days. Except for ER UMa
stars, it was the shortest supercycle ever recorded, leaving SS UMi with
its 84.7 day value behind (however, as was demonstrated by Olech et al.
2006, SS UMi could sometimes switch to normal behavior with much longer
supercycles).

Detailed CCD photometry of BF Ara in superoutburst was done by
Kato et al. (2003) in August 2002. Good coverage of the eruption
allowed for precise determination of the superhump period
which was 0.08797(1) days and seemed to be constant throughout
the entire superoutburst. Other characteristics of the star showed
that it resembles ordinary SU UMa stars rather than active ER UMa
objects.

Table 1 presents the journal of our CCD observations of BF Ara. In
total, we observed the star for almost 31 hours on 15 nights and
collected 666 exposures.

\begin{table}[h!]
\caption{Observational journal for the BF Ara campaign.}
\centering
\begin{tabular}{l c c c c} \hline \hline
Date in 2007    &Start [UT]          &End [UT]   &No. of points  &Duration [h]
\\
\hline
Aug 27           &54339.71157    &54339.80641    &59             &2.28 \\
Aug 28           &54340.71024    &54340.79588    &53             &2.06 \\
Aug 29           &54341.70433    &54341.80002    &60             &3.00 \\
Aug 30           &54342.74061    &54342.80826    &43             &1.62 \\
Aug 31           &54343.70650    &54343.79958    &59             &2.23 \\
Sep 01           &54344.70766    &54344.77470    &39             &1.61 \\
Sep 02           &54345.70788    &54345.79830    &56             &2.17 \\
Sep 03           &54346.70939    &54346.79243    &52             &1.99 \\
Sep 04           &54347.71067    &54347.79506    &44             &2.03 \\
Sep 05           &54348.71216    &54348.79240    &45             &1.93 \\
Sep 06           &54349.71031    &54349.79907    &28             &2.13 \\
Sep 07           &54350.71358    &54350.79352    &30             &1.91 \\
Sep 08           &54351.71438    &54351.79802    &33             &2.01 \\
Sep 09           &54352.71238    &54352.79422    &32             &1.96 \\
Sep 10           &54353.71619    &54353.79464    &33             &1.88 \\
\hline
\label{table1}
\end{tabular}
\end{table}

\section{Light curve}

The light curve of BF Ara from 15 consecutive nights starting on
Aug 27 and ending on Sep 10, 2007 is presented in Fig. 1. Such a
span corresponds to the Rayleigh frequency resolution of $0.07$
c/d. The median separation of observations is $0.0016$ d and thus
the effective Nyquist frequency amounts to 300 c/d. Time sampling
of our observations is non optimal. Large daily gaps and short
nightly runs produced a spectral window with 10 aliases separated by
1 c/d exceeding half power of the central peak. In consequence any
period estimated from our data may suffer from the $1$ c/d ambiguity.
In order to suppress any low frequency long-term irregular
variations night averages were subtracted from the observed
magnitudes.

\clearpage

~

\vspace{10.9cm}

\includegraphics{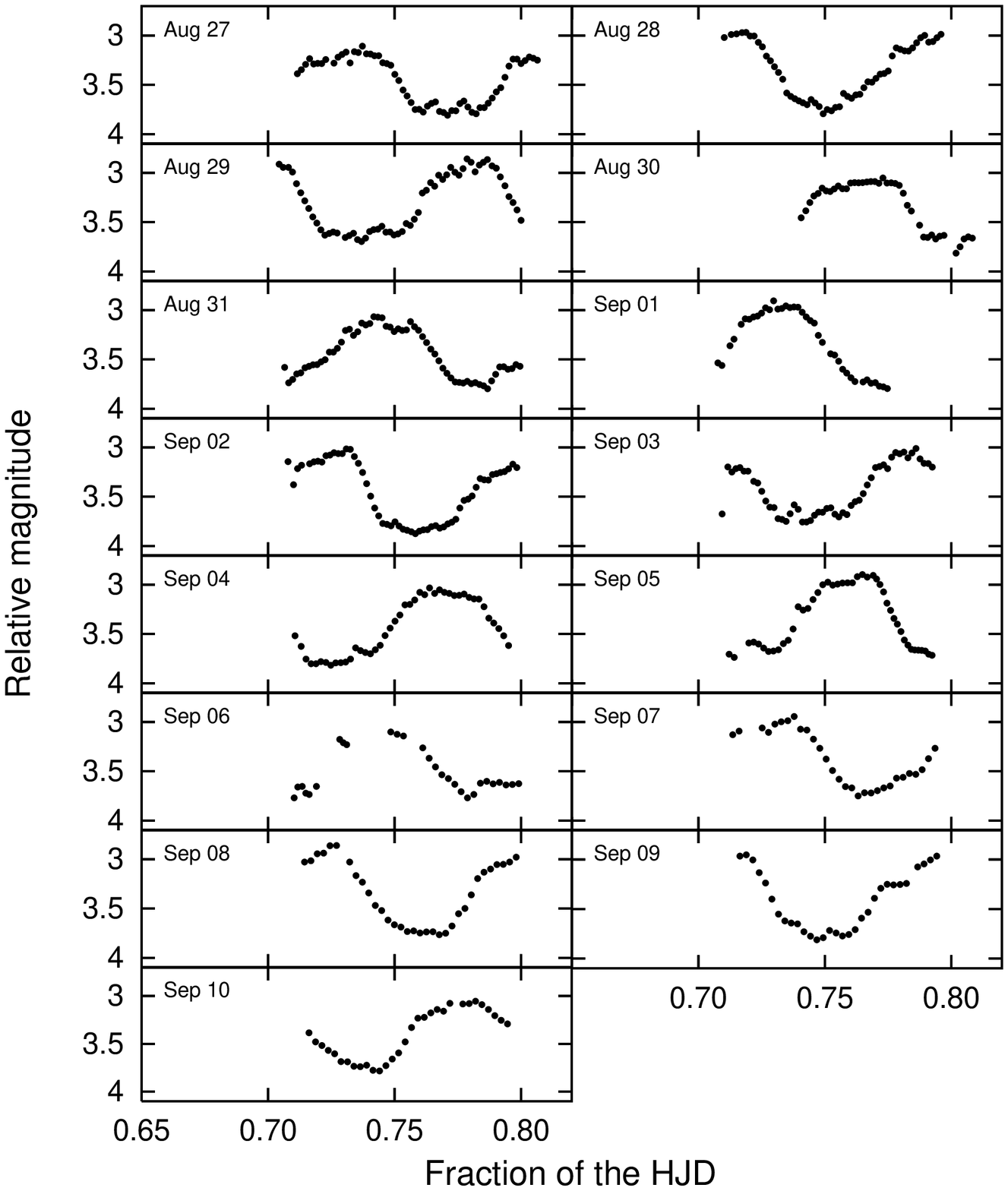}

\begin{figure}[h]
\caption{\sf Light curves of BF Ara obtained on 15 consecutive nights
of August and September of 2007.}
\end{figure}

\subsection{A permanent hump}

Presence of a periodic modulation with peak-to-peak amplitude reaching
up to 1 magnitude was apparent from first glance at the data. The
harmonic analysis-of-variance (ANOVA - Schwarzenberg-Czerny, 1996)
periodogram revealed a pronounced alias pattern centered at the
frequency $f_0\approx 12.173$ c/d (Fig. 2).

\vspace{6.2cm}

\includegraphics{fig2.ps}

\begin{figure}[h]
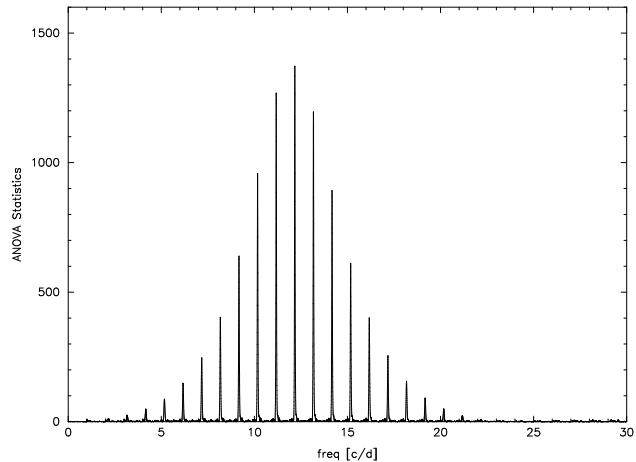

\caption{\sf The harmonic analysis-of-variance (ANOVA) periodogram for
light curves of BF Ara.}
\end{figure}

However, both frequency and amplitude did vary during our observations.
This became first apparent after splitting data in two halves and
analysing them separately. Fitting by least squares of Fourier series of
3 harmonics yields frequencies $12.2013$ and $12.1602$ c/d for the early
and late observations, respectively. The corresponding amplitudes of
first harmonics are 0.79 and 0.68 mag.

In order to study the change of frequency in more detail we assumed
linear dependence of the fundamental frequency and fitted all data with
a Fourier series of 3 harmonics. The best fit yields $f_0=12.1765\pm
0.0011$ c/d and ${\rm df_0/dt}=-0.00529 {\rm c/d}^2$, i.e. $P=0.082125$
d and ${\rm dP/dt}=3.57\cdot10^{-5}$. These values correspond to the
mean epoch of 2454345.8583 HJD. Our errors account for correlation of
3.2 residuals, on the average (Schwarzenberg-Czerny, 1991). In our fit the
$2f_0$ and $3f_0$ harmonics are significant, as they exceed errors 3
times or more. In Fig. 2 we refrained from plotting of the AOV
periodogram for 3 harmonics, as it esthetic is diminished by ghost
sub-harmonics $f_0/2$ and $f_0/3$. However, its closer inspection
reveals that because of extra information from harmonics, its $\pm 1$
c/d aliases are damped compared to Fig. 2. The combined evidence makes
us to believe that for the $f_0$ frequency the 1 c/d ambiguity remains
insignificant. Accounting for the frequency change improved our fit
significantly. The corresponding F statistics rose from 1376 to 2213.
The corresponding shift of phase at time limits of our observations
reaches as much as 0.17 of the period.

\subsection{Secondary modulation}

Prewhitening our data with a constant frequency $f_0$ yields an
unsatisfactory result. Several overlapping alias patterns remain
in the residual periodogram. However, subtraction of the Fourier
series accounting for period change performed well. The
periodogram of residuals becomes clean except for a single alias
pattern centered at $f=12.88$ c/d (Fig. 3). Note that several
alias peaks of comparable height reach values as high as 100. For
a statistician the following conclusion holds: i) the periodic
modulation is detected securely and ii) no unique value of
frequency $f$ may be assigned. No contradiction exists between i)
and ii) as in statistics they correspond to very different
procedures: hypotheses testing and parameter estimation. To be
conservative, we ought to account here for correlation of $2.4$
points on the average (Schwarzenberg-Czerny, 1991). Still, detection
in i) remains significant well above the confidence level of
$0.999$. As far as ii) is concerned, the modulation discussed in
this section corresponds to the signal-to-noise as small as
$S/N=1$ and power in harmonics remains negligible, unlike in Sect.
4.1. In such a situation and for our poor window function the
power in $\pm 1 {\rm c/d}$ aliases almost reaches that of the
central peak, making our frequency determination ambiguous.

\vspace{6.2cm}

\includegraphics{fig3.ps}

\begin{figure}[h]
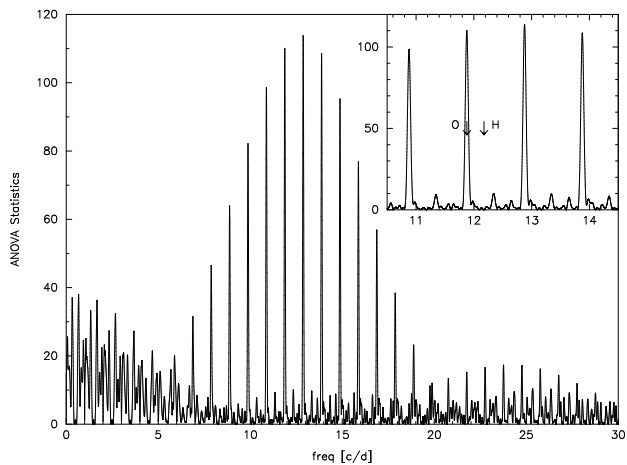

\caption{\sf The harmonic analysis-of-variance (ANOVA) periodogram
for prewhitened light curves of BF Ara. Inset shows close-up of
the frequency pattern in the vicinity of main peak. The arrows
marked $O$ and $H$ indicate respectively the superhump and orbital
frequencies}
\end{figure}

At this point we use the extra information from previous
observations of superhumps in BF Ara (Kato et al. 2003) and
identify $f_1=11.8799\pm 0.0030$ c/d as the true frequency of
the modulation. Its nature will be discussed later. The
corresponding (half)amplitude is $0.072\pm 0.010$ mag. Subtraction
of this latter sinusoid leaves no significant features in the
periodogram up to its effective Nyquist frequency of $300~{\rm
c/d}$.

\section{The O-C diagram}

As we already mentioned, higher amplitude signal was changing its
frequency with time. To check its behaviour we decided to use the $O -
C$ analysis for times of maxima and minima. They were determined by
fitting the polynomials to the observational points in the vicinity of
their extrema. We were able to determine 15 times of maxima and
13 times of minima, which are listed in Table 2 together with associated
errors, cycle numbers $E$, and $O-C$ values.

\begin{table}[!h]
\caption{Times of maxima and minima observed in the light curve of BF Ara
in quiescence.}
\smallskip
\centering
\begin{tabular}{r c c r | r c c r}
\hline \hline
Cycle & ${\rm HJD}_{\rm max}-2 454 300$ & Error &
$O-C$ & ~~Cycle & ${\rm HJD}_{\rm min}-2 454 300$ & Error &
$O-C$\\ $E$ & $[d]$ & $[d]$ & $[cycles]$ & $E$ & $[d]$ & $[d]$
& $[cycles]$ \\ \hline 0 & 39.7340 & 0.001   & 0.037 & 0 & 39.7710
& 0.001 & 0.086\\ 12 & 40.7189 & 0.001 & 0.024 & 12 & 40.7495 &
0.004 & $-0.008$\\ 25 & 41.7860 & 0.002  & 0.011 & 24 & 41.7370 &
0.002 & $0.007$\\ 37 & 42.7690 & 0.001 & $-0.025$ & 37 & 42.8020 &
0.002 & $-0.035$\\ 49 & 43.7563 & 0.001 & $-0.009$ & 49 & 43.7868
& 0.001 & $-0.052$\\ 61 & 44.7380 & 0.002 & $-0.061$ & 73 &
45.7582 & 0.002 & $-0.066$\\ 73 & 45.7309 & 0.003 & 0.023 & 85 &
46.7422 & 0.003 & $-0.093$\\ 86 & 46.7809 & 0.003 & $-0.198$ & 97
& 47.7248 & 0.002 & $-0.137$\\ 98 & 47.7639 & 0.004 & $-0.234$ &
122 & 49.7790 & 0.002 & $-0.143$\\ 110 & 48.7647 & 0.002 &
$-0.054$ & 134 & 50.7633 & 0.002 & $-0.167$\\ 122 & 49.7477 &
0.005 & $-0.090$ & 146 & 51.7677 & 0.001 & $0.054$\\ 134 & 50.7378
& 0.001 & $-0.040$ & 158 & 52.7471 & 0.001 & $-0.029$\\ 146 &
51.7259 & 0.003 & $-0.014$ & 170 & 53.7430 & 0.001 & $0.088$\\ 158
& 52.7190 & 0.001 & 0.073 &  &  & & \\ 171 & 53.7819 & 0.001 &
0.009 &  &  &  & \\ \hline \hline
\end{tabular}
\end{table}
\bigskip

A least-squares linear fit to these data gives the following ephemeris
for the maxima:

\begin{equation}
\textrm{HJD}_\textrm{max} = 2454346.79718(35)+0.082165(6) \cdot (E - E_0)
\end{equation}

\noindent and for the minima:

\begin{equation}
\textrm{HJD}_\textrm{min} = 2454346.83204(39)+0.082187(6) \cdot (E - E_0)
\end{equation}
\smallskip

\noindent where $E_0=86$.

Taking these two determination of period of modulation and the value
obtained from ANOVA statistics, we conclude that mean value of period
was 0.082159(6) days.

\clearpage

~

\vspace{7.5cm}

\includegraphics{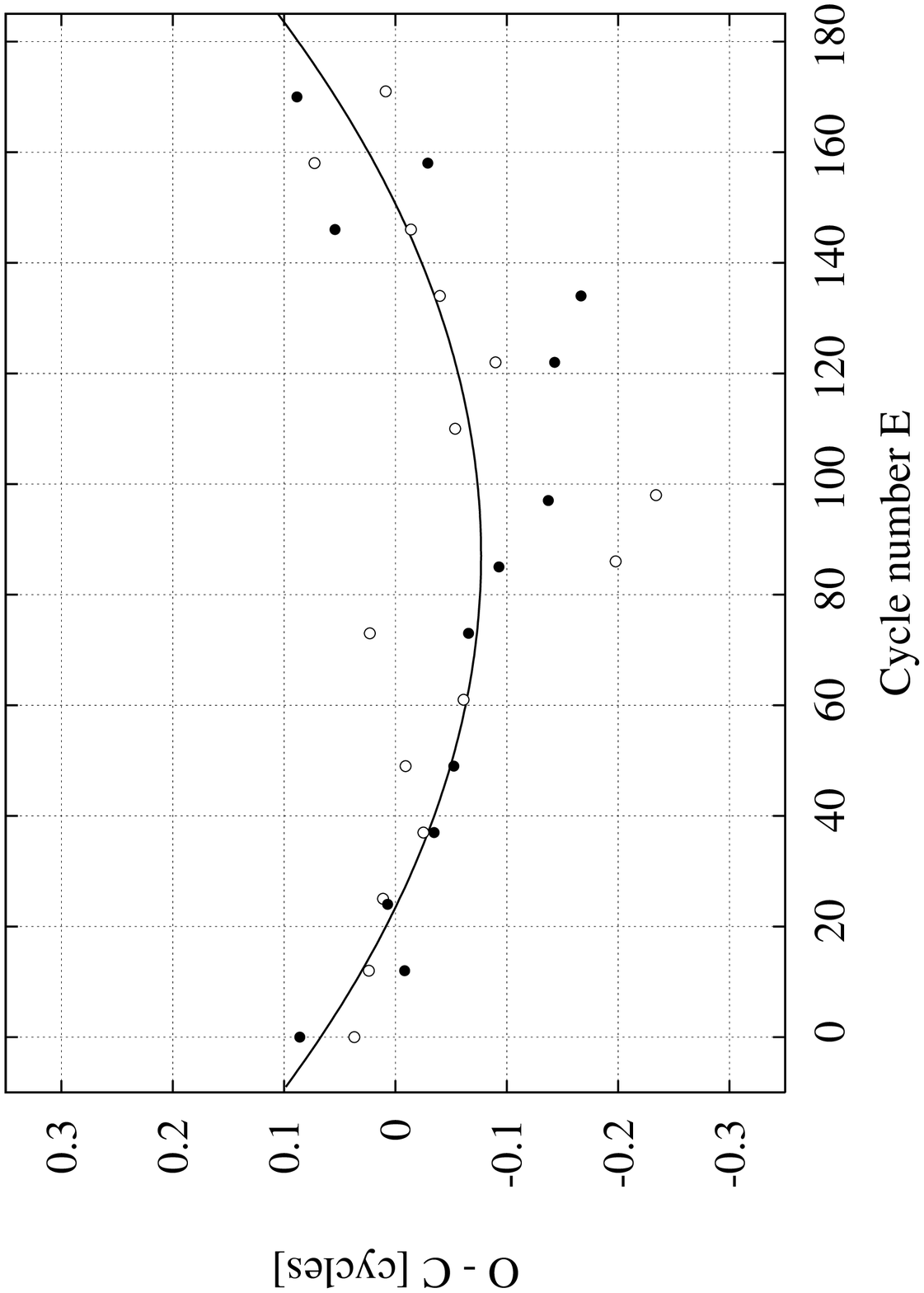}

\begin{figure}[h]
\caption{\sf $O-C$ diagram for the times of maxima (open circles) and
minima (filled circes) observed in the light curve of BF Ara. Solid line
is the fit to the combined $O-C$ values assuming period derivative of
$\dot{P}/P_{sh}=-3.8(3)\times10^{-5}$}
\end{figure}

The $O-C$ values computed according to the ephemeris (1) and (2) are
listed in Table 2 and also shown in Figure 4. It is clear that BF Ara,
during our run, showed a clear change of period. A second-order
polynomial fit the to moments of maxima is expressed by the following
ephemeris:

\begin{equation}
\textrm{HJD}_\textrm{max} =  2454346.79271(68) + 0.082163(6) \cdot (E - E_0) +
1.10(14)\cdot 10^{-6} \cdot (E - E_0) ^2
\end{equation}

\noindent and for the moments of minima:

\begin{equation}
\textrm{HJD}_\textrm{min} =  2454346.82347(74) + 0.082180(6) \cdot (E - E_0) +
2.10(16)\cdot 10^{-6} \cdot (E - E_0)^2
\end{equation}
\smallskip

To use the information both from maxima and minima, we used times and
$O-C$ values of minima, to construct virtual maxima. Then a second-order
polynomial fit was made again resulting in quadratic term of value
equal to $1.56(11)\cdot 10^{-6}$. This corresponds to the period derivative
of $\dot{P}/P_{sh}=-3.8(3)\times10^{-5}$ which is with ideal agreement with
value obtained in Sect. 4.1.

\section{Disscusion}

\subsection{The orbital period}

First, we focus on modulation described in Sect. 4.2. Its
peak-to-peak amplitude is 0.15 mag and one of several acceptable
period values is 0.084176(21) days. The corresponding modulation
appears stable during our 15-days run (see Fig. 5). We proceed to
verify this period value whether it represents
the orbital period of BF Ara.

\clearpage

~

\vspace{5.2cm}

\includegraphics{fig5.ps}

\begin{figure}[h]
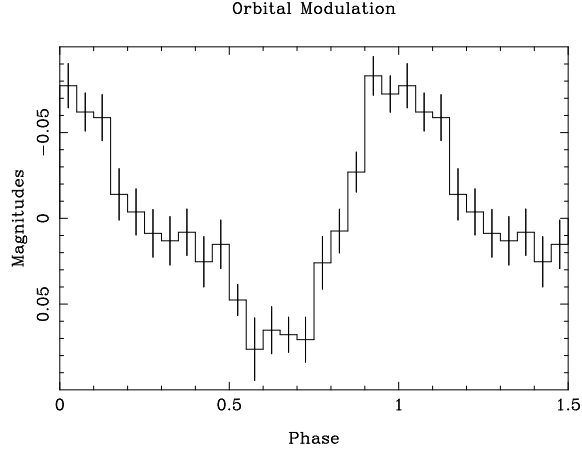

\caption{\sf The mean shape of low amplitude modulation descibed in
Sect. 4.2, after prewhitening with the light variations corresponding
to the high amplitude permanent hump.}
\end{figure}

From their extensive observations of superhumps during superoutburst
Kato et al. (2003)  derived the superhump period $P_{\rm
sh}=0.08797(1)$. It is known, that in SU UMa stars the superhump and
orbital periods, $P_{\rm sh}$ and $P_o$, obey the Stolz and Schoembs
(1984) relation. For this purpose one defines the superhump period
excess as $\epsilon = (P_{\rm sh}-P_o)/P_{o}$. In Fig.~6 we present a
plot of superhump excess against the superhump period for all CVs with
known orbital periods. For BF Ara we considered periods of several
aliases for the secondary modulation of Sect~4.2 as tenative orbital
periods and calculated the corresponding superhump excesses. All but one
value obtained in this way deviated widely from the relation in
Fig.~6. The only consistent value, $\varepsilon=4.51\% \pm 0.03\%$
corresponds to our proposed orbital period of BF Ara and is plotted as
filled square. It is clear that BF Ara follows this relation perfectly,
giving credit to our belief that the secondary small amplitude
modulation reflects the orbital wave. It is also worth to mention that
its value of $P_{\rm orb}=121.2$ minutes locates BF Ara in the period gap.

\vspace{7.5cm}

\includegraphics{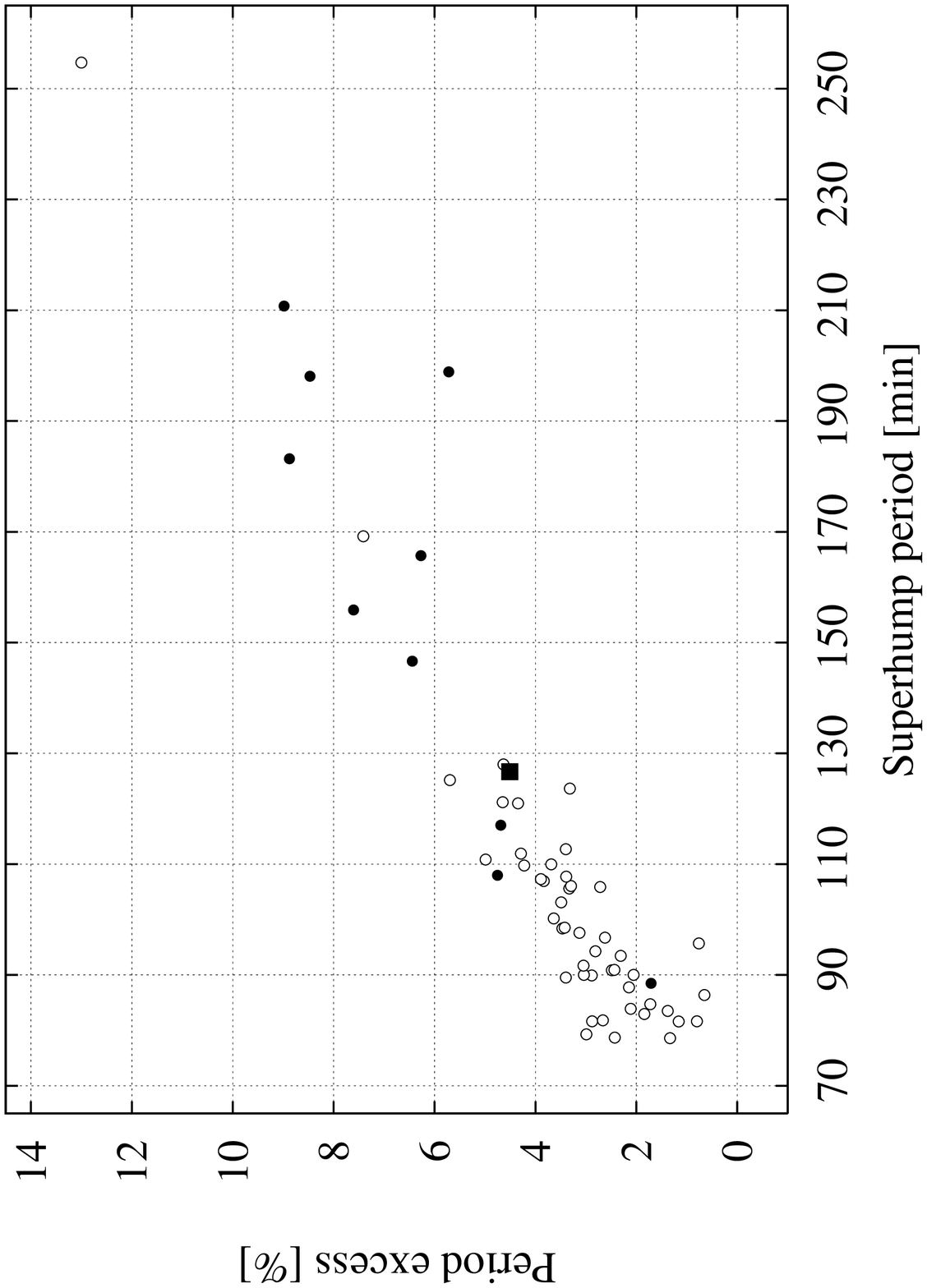}

\begin{figure}[h]
\caption{\sf The relation between superhump period and period excess for
ordinary SU UMa stars (open circles), objects other than dwarf novae
(filled circles) and for BF Ara (filled square).}
\end{figure}
\smallskip

It has been established that the period excess in SU UMa stars
correlates with the mass ratio $q=M_2/M_1$. The corresponding relation
may be approximated in the following way (Osaki 1985):

\begin{equation}
\epsilon\approx\frac{0.23q}{1+0.27q}
\end{equation}
\medskip

\noindent Thus the known period excess $\epsilon$ of BF Ara can be
used to estimate its mass ratio as $q\approx 0.21$.

Finaly, we decided to give the ephemeris for maxima and minima of
orbital modulation. They are as follows:

\begin{equation}
\textrm{HJD}_\textrm{max} = 2454345.8180(18) + P_{\rm orb} \cdot N
\end{equation}

\begin{equation}
\textrm{HJD}_\textrm{min} = 2454345.7886(18) + P_{\rm orb} \cdot N
\end{equation}
\smallskip

\noindent The moments od maxima and minima were determined by fitting
data from Fig. 5 with a Fourier series up to $2f_{\rm orb}$ and then
finding its extrema.

\subsection{The negative superhumps}

Now we verify whether the permanent hump modulation discussed in Sect.
4.1 fits the negative hump pattern. Its mean period was 0.082159(4)
days, increasing at the rate  of $\dot P/P_{\rm h} = 3.8(3)\cdot
10^{-5}$. The mean amplitude of this modulation was 0.7 mag and its mean
shape is shown in Fig. 7.

\vspace{5.6cm}

\includegraphics{fig7.ps}

\begin{figure}[h]
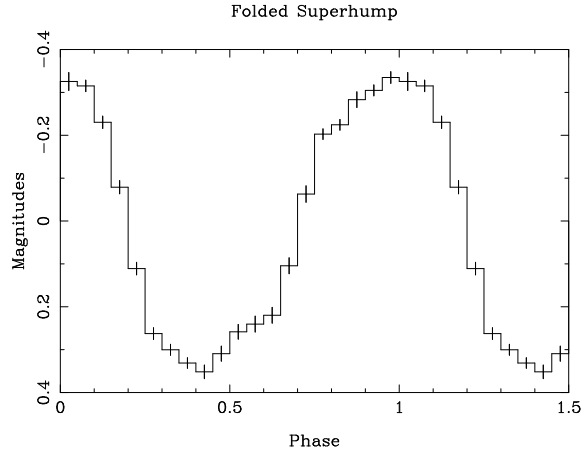

\caption{\sf The mean shape of the negative superhump descibed in
Sect. 4.1, after prewhitening with the light variations corresponding
to the orbital hump.}
\end{figure}

There are cataclysmic variables exhibiting three close periods of
modulation: the orbital wave, ordinary superhumps and, additionally, so
called negative superhumps. The periods of ordinary and negative humps,
$P_{\rm sh}$ and $P_{\rm nh}$ are respectively longer and shorter than the
orbital period $P_{\rm orb}$. It is believed that negative superhumps
arise due to  the classical precession of a tilted accretion
disk. According to Wood and Burke (2007) in these stars the accretion
disk is tilted out of the orbital plane. The negative hump modulation is
caused by migration of the hot spot across the face of the disk. Without
any tilt, the stream of accreted matter hits the edge of the disk
roughly at the fixed distance from white dwarf in the orbital plane. For
a tilted disc this situation happens only when the hot spot crosses the
nodes of a disk. At other orbital phases stream misses outer parts of the
disc, penetrating deeper in the gravitational potential well of the
white dwarf. This in turn causes brightening of the bright spot. Since
only one face of the optically thick disk is visible, only one
brightening per orbit is observed. A slow retrograde precession of the
tilted disk results in a period slightly shorter than the orbital
period.

Positive and negative superhumps are observed in variety of objects.
Among them are double-degenerate systems such as AM CVn (Skillman et al.
1999), ordinary SU UMa stars like V503 Cyg (Harvey et al. 1995),
classical novae like V1974 Cyg (Semeniuk et al. 1995, Retter, Leibowitz
and Ofek 1997), X-ray systems containing neutron stars like V1405 Aql
(Retter et al. 2002a) and intermediate polars like TV Col (Retter et al.
2002b). It is interesting that the super- and negative-hump periods in
all  objects seem to follow a unique relation. The respective period
excess and defect are defined as $\varepsilon=(P_{\rm sh}-P_{\rm
orb})/P_{\rm orb}$ and $\varepsilon_-=(P_{nh}-P_{\rm orb})/P_{\rm orb}$.
In Fig.~8 taken from Retter et al. (2002a) we plot the defect and excess
ratio $\phi=\varepsilon_-/\varepsilon$ against the orbital period. The
picture suggests a unique relation.

From the observed the hump and orbital periods we determine the
period defect of BF Ara. It amounts to $\epsilon_- = -2.44\% \pm
0.02\%$ and yields the ratio of the positive and negative hump
defect and excess $\phi = -0.540\pm0.006$. We added the
corresponding point in Fig.~8. It fits very well the overall
trend, adding credit to our interpretation of this modulation as
the negative hump.

\vspace{8.9cm}

\includegraphics{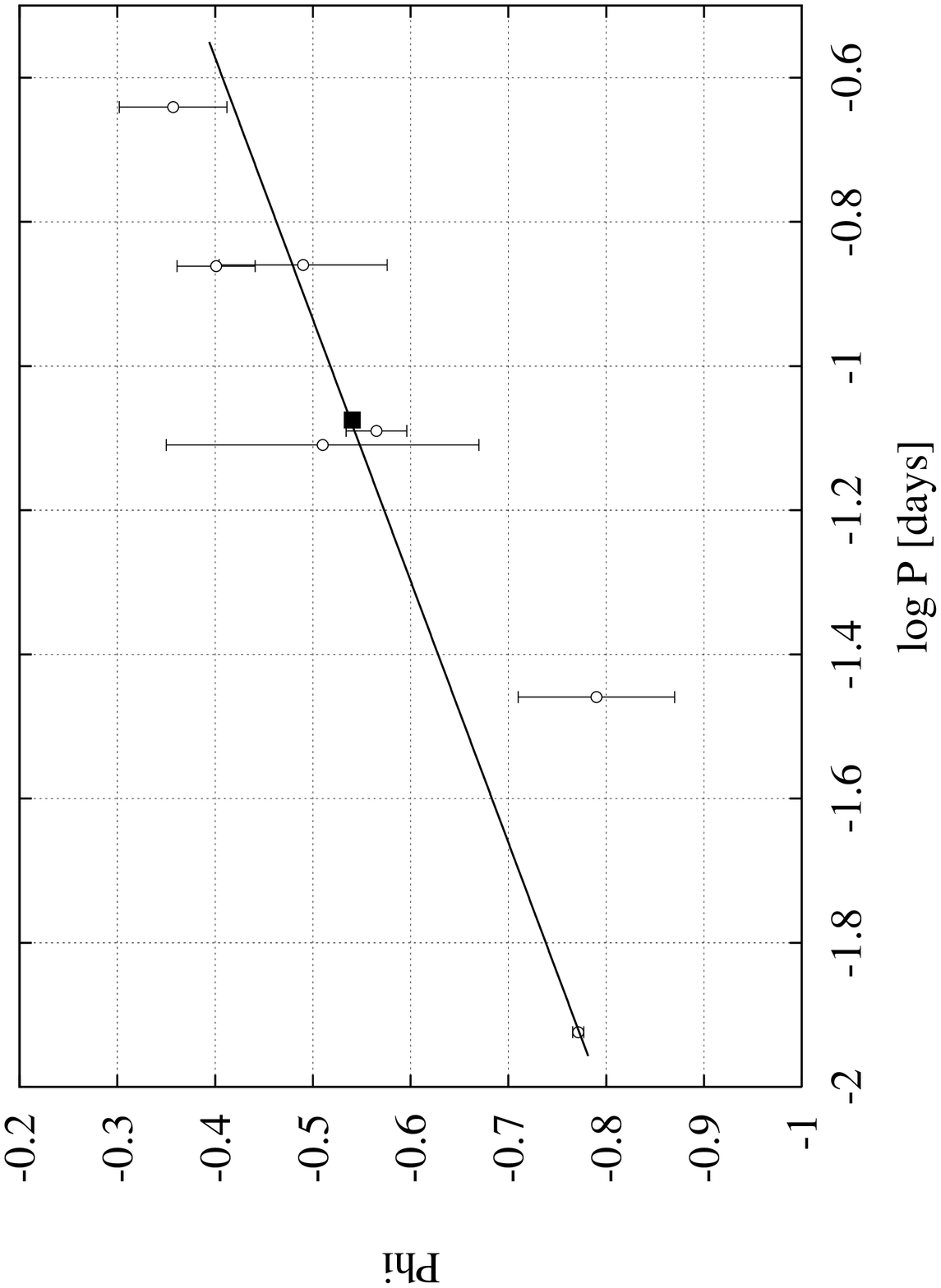}

\begin{figure}[h]
\caption{\sf The relation between the ratio of negative
and positive period excess and orbital period for cataclysmic variables
showing both positive and negative superhumps. BF Ara is plotted with
solid rectangle. Error of this determination is of the size of the symbol.
Straight line corresponds to the fit given in equation (6).}

\end{figure}
\smallskip

Fitting a straight line to the points corresponding to
individual variables one obtains the following relation:

\begin{equation}
\phi= 0.276(9) \cdot \log P - 0.232(14)
\end{equation}

\section{Summary}

Our main conclusions may be summarized as follows:

\begin{itemize}

\item Light curve of BF Ara in quiescence is dominated by high amplitude
signal with mean period of 0.082159(4) days which during two weeks
observing run increased at the rate of $\dot P/P_{\rm sh} =
3.8(3)\cdot 10^{-5}$

\item We interpret this periodicity as being due to a negative
superhump. Such an interpretation is confirmed by the location of BF Ara
in the Ritter et al. (2002a) relation,

\item BF Ara seems to be a twin of V503 Cyg (Harvey et al., 1995) -
both stars are very active (supercycles of 80--90 days), both show
large amplitude negative superhumps in quiescence characterized by
changing value period,

\item Prewhitening of the original light curve with the main
periodicity resulted in the discovery of another modulation
with constant period equal to 0.084176(21) days, which is
interpreted as the orbital period of the system. This value makes BF
Ara another in-the-gap cataclysmic variable,

\item Knowing the ordinary superhump period measured by Kato et al. (2003)
we were able to calculate the period excess as equal to $4.51\% \pm 0.03\%$
which indicates the mass ratio of $q\approx0.21$.

\end{itemize}

\bigskip \noindent {\bf Acknowledgments.} ~We acknowledge generous
allocation of  the SAAO 1-m telescope time. We would like to thank
Prof. J\'ozef Smak for fruitful discussions. This work was supported
for SALT grant number 76/E-60/SPB/MSN/P-03/DWM 35/2005-2007.

\end{document}